\documentclass[twocolumn,prc,showpacs,showkeys]{revtex4}
\usepackage{amsfonts}
\usepackage{amsmath}
\usepackage{amssymb}
\usepackage{graphicx}
\usepackage{rotating}

\providecommand{\U}[1]{\protect\rule{.1in}{.1in}}
\providecommand{\U}[1]{\protect\rule{.1in}{.1in}}
\providecommand{\U}[1]{\protect\rule{.1in}{.1in}}

\begin{document}

\title{Photo-induced nuclear cooperation\\
}
\author{P\'{e}ter K\'{a}lm\'{a}n}
\author{Tam\'{a}s Keszthelyi}
\affiliation{Budapest University of Technology and Economics, Institute of Physics,
Budafoki \'{u}t 8. F., H-1521 Budapest, Hungary\ }

\begin{abstract}
Reactions $n_{\gamma }\gamma +$ $d+$ $_{z_{2}}^{A_{2}}X\rightarrow \left(
n_{\gamma }\pm 1\right) \gamma +$ $p+$ $_{z_{2}}^{A_{2}+1}X$ and especially
the reaction $d+$ $_{z_{2}}^{A_{2}}X\rightarrow \gamma +$ $p+$ $%
_{z_{2}}^{A_{2}+1}X$, called photo-induced nuclear cooperation and
cooperative spontaneous $\gamma $ emission with neutron exchange,
respectively, are investigated theoretically. In the case of photo-induced
nuclear cooperation it is supposed that the energy of $\gamma $ photons of
the beam is less than the binding energy of the deuteron. The cross section
and the transition probability per unit time, respectively, are determined
with the aid of standard second order perturbation calculation of quantum
mechanics. The calculations are extended to photo-induced nuclear
cooperation and cooperative spontaneous $\gamma $ emission with proton
exchange as well. With the aid of the results obtained, recent observations
of nuclear activity of samples of large deuteron content after irradiation
by photon-flux of photon energy smaller than the deuteron binding energy are
discussed.
\end{abstract}

\startpage{1}
\endpage{}
\maketitle

\section{Introduction}

In a recent report \cite{steinetz} observation of nuclear activity in
deuterated materials subjected to low-energy photon beam was announced. In
the experiments different mixtures of deuterated materials were subjected to
a photon beam of photon energy less than the deuteron binding energy. The
specimens are made from $ErD_{2}+C_{36}D_{74}+Mo$ and $%
HfD_{2}+C_{36}D_{74}+Mo$ mixtures. The gamma activity measurements made
after irradiation showed significant presence of $^{163}Er$, $^{171}Er$, $%
^{99}Mo$ and $^{101}Mo$ radioisotopes of zero natural abundance in specimens
of deuterated erbium and $^{180m}Hf$, $^{181}Hf$, $^{99}Mo$ and $^{101}Mo$
radioisotopes also of zero natural abundance in specimens of deuterated
hafnium. In both cases presence of $^{99m}Tc$ and $^{101}Tc$ isotopes and
creation of neutrons were also found. In control experiments, i.e. with
specimens made from hydrogenated or non-gas-loaded (without any hydrogen
isotope) materials gamma spectra revealed no new isotopes. In this work it
is attempted to give a mechanism to expound the observations.

The results of \cite{steinetz} indicate that to induce nuclear activity a
joint presence of deuterons and a photon flux is necessary. Thus it is
expected that those processes may be responsible for nuclear activity in
which electromagnetic radiation virtually breaks up deuterons and than the
virtually-free neutron is captured by an other nucleus. Namely the reaction%
\begin{equation}
n_{\gamma }\gamma +\text{ }d+\text{ }_{z_{2}}^{A_{2}}X\rightarrow \left(
n_{\gamma }\pm 1\right) \gamma +\text{ }p+\text{ }_{z_{2}}^{A_{2}+1}X+D_{j},
\label{R1}
\end{equation}%
which further on will be called photo-induced nuclear cooperation with
neutron exchange, will be investigated. Here $\gamma $ denotes photon of
energy less than the binding energy $B=2.225$ $MeV$ of the deuteron, $%
n_{\gamma }$ is the number of $\gamma $ photons initially present, $d$ is
deuteron which is 'virtually' broken up due to electromagnetic interaction, $%
_{z_{2}}^{A_{2}}X$ stands for the cooperating (target) nucleus which absorbs
the 'virtual' free neutron, $_{z_{2}}^{A_{2}+1}X$ denotes final nucleus (of $%
A_{3}=A_{2}+1$) and $D_{j}$ is the energy of the reaction.

A special case of $\left( \ref{R1}\right) $ when $n_{\gamma }=0$, i.e. when
initially photons are not present. In this case $\left( \ref{R1}\right) $
reads as%
\begin{equation}
d+_{z_{2}}^{A_{2}}X\rightarrow \gamma +p+_{z_{2}}^{A_{2}+1}X+D_{j}.
\label{R5}
\end{equation}%
This reaction is called cooperative spontaneous $\gamma $ emission with
neutron exchange.

The transition probability per unit time and the cross section of processes $%
\left( \ref{R1}\right) $ and $\left( \ref{R5}\right) $ can be determined
with the aid of second order standard perturbation calculation of quantum
mechanics \cite{Landau}. (The term 'virtual' refers to the intermediate
state in standard second order perturbation calculation.)

Accordingly, in photo-induced nuclear cooperative process the
electromagnetic field-matter and strong interactions are essential. The
interaction Hamiltonian $H_{I}$ has two terms:%
\begin{equation}
H_{I}=V^{em}+V^{st}.  \label{HI}
\end{equation}%
$V^{em}$ describes the interaction of electromagnetic radiation with matter
primarily with deuteron and $V^{st}$ stands for strong interaction acting
between the 'virtual' free neutron and the cooperating (target) nucleus $%
_{z_{2}}^{A_{2}}X$\ which absorbs the 'virtual' free neutron. In the case of
photo-induced nuclear cooperation 'first' $V^{em}$ and 'after' $V^{st}$ acts
when determining the transition probability per unit time $\left(
W_{fi,\gamma n}\right) $ and the cross section $\left( \sigma _{\gamma
n}\right) $ of this second order process in standard manner (e.g. see \cite%
{Landau}). (The terminology 'first' and 'after' corresponds to the time
ordering of the operators in the calculation.) Since in control experiments 
\cite{steinetz} nuclear activity did not appear, the observed phenomenon may
be attached to 'virtual' photo-disintegration of deuteron. Therefore our
investigation is restricted to this case only and in the description one can
apply the theoretical results obtained during cross section calculation of
real photo-disintegration of the deuteron \cite{Bethe}, \cite{Blatt}. The
momentum of the photon is neglected compared to the momenta of the proton
and the final nucleus $_{z_{2}}^{A_{2}+1}X$.

The photo-induced nuclear cooperation with neutron exchange is dealt with in
Section II, where initial, intermediate and final states and energy
relations of the process, the cooperation factor, the transition probability
per unit time of spontaneous photo-induced nuclear cooperation and the cross
section of photo-induced nuclear cooperation with neutron exchange are
given. Section III. is devoted to nuclear cooperation with proton exchange
dealing with the Coulomb factor in nuclear cooperation with proton exchange,
the transition probability per unit time of spontaneous decay and the cross
section of photo-induced nuclear cooperation with proton exchange. In
section IV. the explanation of observations is discussed comparing the rate
of the process induced by irradiation of a photon flux and the rate of the
spontaneous process. As a numerical example the rate of unstable $^{99}Mo$
isotope creation in the spontaneous process is given too. Section V. is
devoted to conclusions. In the Appendix the interaction Hamiltonians, the
matrix elements of $V^{em}$ and $V^{st}$, which are necessary to second
order perturbation calculation, and some details of calculation of $%
W_{fi,\gamma n}$ are given.

\section{Photo-induced nuclear cooperation with neutron exchange}

\subsection{Initial, intermediate and final states}

Deuterons are somewhere in the sample of volume $V_{s}$. Their initial state
describing the motion of the center of mass $\left( CM\right) $ of the
deuteron is a wave of amplitude $V_{s}^{-1/2}$. This is the most simple
choice. The state of the deuteron in the relative (neutron-proton
separation) coordinate $\left( \mathbf{r}\right) $\ reads as $\varphi
_{d}=\left( 4\pi \right) ^{-1/2}\sqrt{\alpha /(2\pi )}e^{-\alpha r}/r$,
where $\alpha =\hbar ^{-1}\sqrt{Bm_{0}}$ and $r=\left\vert \mathbf{r}%
\right\vert $ \cite{Bethe}, \cite{Blatt}.

The initial state of target nucleus (of mass number $A_{2}$) is a wave of
amplitude $V_{t}^{-1/2}$, i.e. the target nucleus is somewhere in the volume
of normalization $V_{t}$.

The motion of center of mass of intermediate neutron and the final proton
states are plane waves of wave number vector $\mathbf{k}_{n}$, $\mathbf{k}%
_{p}$ and volumes of normalization $V_{n}$, $V_{p}$, respectively. The
cooperation (of deuteron) by neutron with an other nucleus is taken into
account with the aid of spherical waves the source of which is the deuteron
and which behaves far away (at incidence on nucleus $_{z_{2}}^{A_{2}}X$) as
a plane wave.

For the final bound neutron states of excitation energy $\varepsilon _{j}$
of nucleus of mass number $A_{3}=A_{2}+1$, where $A_{2}$ is the mass number
of target nucleus, we take $\Phi \left( \mathbf{r}_{n}\right) =\sqrt{%
3/R_{3}^{3}}\phi _{j}\left( R_{3}x\right) Y_{l_{f}m_{f}}\left( \Omega
_{n}\right) $ where $x=r_{n}/R_{3}$, $R_{3}=r_{0}A_{3}^{1/3}$is the radius
of a nucleus of nucleon number $A_{3}$ with $r_{0}=1.2\times 10^{-13}$ $cm$,
the $Y_{l_{f}m_{f}}\left( \Omega _{n}\right) $ is a spherical harmonics and $%
\int_{0}^{\infty }\left\vert \phi _{j}\left( R_{3}x\right) \right\vert
^{2}x^{2}dx=1/3$. In the Weisskopf-approximation to be used $\phi
_{j}^{W}\left( R_{3}x\right) =1$, if $x\leq 1$ and $\phi _{j}^{W}\left(
R_{3}x\right) =0$ for $x>1$, The final state which describes the motion of
center of mass of the final nucleus (of mass number $A_{3}$) is also a plane
wave of wave number vector $\mathbf{k}_{3}$ and of volume of normalization $%
V_{3}$. It is supposed that $V_{3}=V_{t}$.

The wave number (momentum) of the photon is much less than the wave numbers
of the proton and the final nucleus $_{z_{2}}^{A_{2}+1}X$, therefore it is
neglected in the calculation of momentum conservation.

\subsection{Energy relations}

$D_{0n}=\Delta _{-}+\Delta _{+}$ is the energy of reaction into the ground
state of the final nucleus with $\Delta _{-}=\Delta _{d}-\Delta _{p}$ and $%
\Delta _{+}=\Delta _{A_{2}}-\Delta _{A_{2}+1}$ $\left( D_{0n}=\Delta
_{d}-\Delta _{p}+\Delta _{A_{2}}-\Delta _{A_{2}+1}\right) $. $\Delta
_{A_{2}} $, $\Delta _{A_{2}+1}$, $\Delta _{d}$, $\Delta _{p}$ and $\Delta
_{n}$\ are the energy excesses of neutral atoms of mass numbers $A_{2}$, $%
A_{2}+1$, deuteron, proton and neutron, respectively \cite{Shir}. It is
possible (energetically allowed) that the final nucleus is created in an
excited state of energy $\varepsilon _{j}(>0)$ above its ground state. The
reaction energy $D_{j}=D_{0n}-\varepsilon _{j}$ belongs to that reaction
which has final state of excitation energy $\varepsilon _{j}$. It is useful
to introduce the quantity 
\begin{equation}
\Delta _{j}^{\pm }=D_{0n}-\varepsilon _{j}\pm E_{\gamma }.  \label{DeltaJ}
\end{equation}%
Here $E_{\gamma }$ is the energy of the photon. $\Delta _{j}^{\pm }$ is the
energy which is shared between the kinetic energies of final nucleus and
proton. The upper $+$ and $-$ signs throughout correspond to absorption and
emission of photon.

\subsection{Spontaneous decay by photo-induced nuclear cooperation with
neutron exchange}

In the case of $n_{\gamma }=0$, i.e. initially photons are not present, only
the $V_{E,ki}^{em-}$ and $V_{M,ki}^{em-}$ matrix elements with $g^{-}=1$
give contribution (see Appendix B.). It is the case of cooperative
spontaneous $\gamma $ emission (see $\left( \ref{R5}\right) $). The phase
space of the emitted photon is $\left( 2\pi \hbar c\right) ^{-3}V_{\gamma
}E_{\gamma }^{2}d^{3}E_{\gamma }d\Omega _{\gamma }$ with which the
expression of transition probability per unit time must be multiplied too. ($%
\hbar $ is the reduced Planck constant and $c$ is the velocity of light.)

\subsubsection{Cooperation factor}

Determining the full transition probability per unit time $W_{fi,\gamma n}$,
the contributions coming from all cooperating nuclei located at a distance $%
L $ far from each other in the case of every possible $L$ value must be
taken into account. The number of cooperating nuclei in a shell of sphere of
radius $L$ and width $dL$ reads as $4\pi L^{2}dLn_{A_{2}}$ with $n_{A_{2}}$
the number density of nuclei of nuclear number $A_{2}$. The emitted
'virtual' neutron wave has an amplitude $A(L)$ in this shell (see Appendix
C). Using $\left\vert A(L)\right\vert =\sin \left( k_{n}L\right) /\left(
k_{n}L\right) $ or $\left\vert A(L)\right\vert =\cos \left( k_{n}L\right)
/\left( k_{n}L\right) $ it results a factor 
\begin{equation}
F_{coop}=\int_{L_{0}}^{\Lambda }A^{2}\left( L\right) 4\pi
L^{2}dLn_{A_{2}}=k_{n}^{-2}2\pi \Lambda n_{A_{2}},  \label{Fcoop}
\end{equation}%
which is responsible for cooperation and it is called cooperation factor
further on, in $W_{fi,\gamma n}$, where small terms ($k_{n}^{-2}2\pi
L_{0}n_{A_{2}}$ and $\pm k_{n}^{-3}\pi n_{A_{2}}\left[ \sin \left(
2k_{n}\Lambda \right) -\sin \left( 2k_{n}L_{0}\right) \right] )$ are
neglected since $L_{0}\ll \Lambda $ and $k_{n}^{-1}\ll \Lambda $. Here $%
\Lambda $ is the cooperation length and $L_{0}$ is the distance between
deuteron and nearest nucleus of nuclear number $A_{2}$. $\Lambda $ has the
order of magnitude of characteristic linear size $V_{s}^{1/3}$ of the
sample. $\Lambda =V_{s}^{1/3}$ is supposed further on.

However, each neutron may contribute to the effect and $W_{fi,\gamma n}$
must be multiplied by their number $N_{d}=V_{s}n_{d}$ with $n_{d}$ the
number density of deuterons.

\subsubsection{Transition probability per unit time of spontaneous
photo-induced nuclear cooperation with neutron exchange}

The full transition probability per unit time $W_{fi,\gamma n}^{-}$ of the
spontaneous process has the form%
\begin{equation}
W_{fi,\gamma n}^{-}=n_{d}\Lambda
n_{z_{2}}r_{A_{2}}A_{3}C_{W}\sum_{\varepsilon _{j}}I_{jn}^{-}  \label{Wfign-}
\end{equation}%
%
%.
Here $n_{z_{2}}$ is the number density of element of charge number $z_{2}$, $%
r_{A_{2}}$ is the natural abundance of isotope of mass number $A_{2}$ and%
\begin{equation}
C_{W}=\frac{6\alpha _{f}r_{0}^{3}c}{\pi ^{2}\left( 1-\alpha r_{t}\right) }%
\frac{B^{2}V_{0}^{2}m_{0}c^{2}A_{3}}{\alpha ^{6}\left( \hbar c\right)
^{5}\left( A_{3}+1\right) }.  \label{CW}
\end{equation}%
Here $\alpha _{f}$ is the fine structure constant ($e^{2}=\alpha _{f}\hbar c$%
), $V_{0}=45$ $MeV$, $m_{0}c^{2}=931,494$ $MeV$ is the atomic mass unit, $e$
is the elementary charge, $r_{t}=1.759\times 10^{-13}$ $cm$ \thinspace\ \cite%
{Arenhovel}, $C_{W}=8.02\times 10^{-45}$ $cm^{5}s^{-1}$, and 
\begin{equation}
I_{jn}^{-}=\int_{0}^{\delta _{j}}u^{3}\left[ F_{jl_{f}}\left( \xi
_{j}^{-}\right) \left[ S^{E}\left( \xi _{j}^{-}\right) +\eta
_{0}^{2}S^{M}\left( \xi _{j}^{-}\right) \right] \right] du  \label{Ij-}
\end{equation}%
with $u=E_{\gamma }/B$, $\delta _{j}=\left( D_{0n}-\varepsilon _{j}\right)
/B $, $\eta _{0}^{2}=0.3684$ and 
\begin{equation}
\xi _{j}^{\pm }=\left( \frac{k_{0j}^{\pm }}{\alpha }\right) ^{2}=\left( 
\frac{A_{3}}{A_{3}+1}\right) 2\left( \delta _{j}\pm u\right) .
\label{k0peralf}
\end{equation}%
Here $\xi _{j}^{-}$ is used. 
\begin{equation}
S^{E}\left( \xi _{j}^{\pm }\right) =\frac{\frac{2}{3}\xi _{j}^{\pm }}{\left[
1+\xi _{j}^{\pm }\right] ^{4}},  \label{SE}
\end{equation}%
\begin{equation}
S^{M}\left( \xi _{j}^{\pm }\right) =\frac{1}{\left[ 1+\xi _{j}^{\pm }\right]
^{2}\left( 1+\xi _{j}^{\pm }\alpha ^{2}a_{s}^{2}\right) },  \label{SM}
\end{equation}%
where $\alpha ^{2}a_{s}^{2}=29.89$,%
\begin{equation}
F_{jl_{f}}\left( \xi _{j}^{\pm }\right) =\frac{\left( 2l_{f}+1\right)
H_{jl_{f}}^{2}(\alpha R_{3}\sqrt{\xi _{j}^{\pm }})}{\left[ \xi _{j}^{\pm
}+1\mp u\right] ^{2}\sqrt{\xi _{j}^{\pm }}}.  \label{Fk0}
\end{equation}%
For the definition of $H_{jl_{f}}$ see $\left( \ref{HknRA3}\right) $, it is
given by $\left( \ref{HknRA3W}\right) $ and $\left( \ref{HknRA3LWA}\right) $
in the Weisskopf- and Weisskopf-long wavelength approximations.

\subsection{Cross section of photo-induced nuclear cooperation with neutron
exchange}

The cross section $\sigma _{\gamma n}^{\pm }$ of photo-induced nuclear
cooperation with neutron exchange due to all cooperating nuclei located in
the sample can be obtained from the transition probability per unit time $%
W_{fi,\gamma n}$ omitting from it the phase space of $\gamma $ and dividing
it by the photon flux $cn_{\gamma }/V_{\gamma }$ \ where the $n_{\gamma
}+1\simeq n_{\gamma }$ approximation is used.

The photo-induced cross section $\sigma _{\gamma n}^{\pm }$ has the form%
\begin{equation}
\sigma _{\gamma n}^{\pm }=\sum_{\varepsilon _{j}}K_{\sigma
}uF_{jl_{f}}\left( \xi _{j}^{\pm }\right) \left[ S^{E}\left( \xi _{j}^{\pm
}\right) +\eta _{0}^{2}S^{M}\left( \xi _{j}^{\pm }\right) \right]
\label{sigmanE}
\end{equation}%
where $S^{E}\left( \xi _{j}^{\pm }\right) $ and $S^{M}\left( \xi _{j}^{\pm
}\right) $ come from contributions due to electric and magnetic parts of
deuteron-photon electromagnetic interaction [see $\left( \ref{SE}\right) $
and $\left( \ref{SM}\right) $], $F_{jl_{f}}\left( \xi _{j}^{\pm }\right) $
is determined by $\left( \ref{Fk0}\right) $, $\xi _{j}^{\pm }$ is given by $%
\left( \ref{k0peralf}\right) $, $u=E_{\gamma }/B$\ and%
\begin{equation}
K_{\sigma }=V_{s}^{-1}\Lambda n_{z_{2}}r_{A_{2}}A_{3}C_{\sigma },
\label{Kg2}
\end{equation}%
with%
\begin{equation}
C_{\sigma }=\frac{6\alpha _{f}r_{0}^{3}}{\left( 1-\alpha r_{t}\right) }\frac{%
V_{0}^{2}m_{0}c^{2}A_{3}}{B\alpha ^{6}\left( \hbar c\right) ^{2}\left(
A_{3}+1\right) },  \label{C0}
\end{equation}%
($C_{\sigma }=1.840\times 10^{-87}$ $cm^{7}$).

\section{Nuclear cooperation with proton exchange}

If the deuteron is 'virtually' splitted up by a photon then the reaction 
\begin{equation}
n_{\gamma }\gamma +\text{ }d+\text{ }_{z_{2}}^{A_{2}}X\rightarrow \text{ }%
\left( n_{\gamma }\pm 1\right) \gamma +n+\text{ }%
_{z_{2}+1}^{A_{2}+1}Y+D_{jp},  \label{R2}
\end{equation}%
which is nuclear cooperation with proton exchange and the reaction 
\begin{equation}
d+\text{ }_{z_{2}}^{A_{2}}X\rightarrow \gamma +\text{ }n+\text{ }%
_{z_{2}+1}^{A_{2}+1}Y+D_{jp},  \label{R4}
\end{equation}%
which is cooperative spontaneuos $\gamma $ emission with proton exchange,
may happen too. Now $D_{jp}=D_{0p}-\varepsilon _{j}$ with $D_{0p}=\Delta
_{d}-\Delta _{n}+\Delta _{A_{2}}-\Delta _{A_{2}+1}$. ($\Delta _{j}^{\pm
}=D_{0p}-\varepsilon _{j}\pm E_{\gamma }$ and $\delta _{j}=\left(
D_{0p}-\varepsilon _{j}\right) /B$). However these reactions are hindered by
the Coulomb repulsion between the proton and the nucleus $_{z_{2}}^{A_{2}}X$%
, which is manifested in the appearance of the Coulomb factor in the matrix
element of $V^{st}$.

\subsection{Coulomb factor in nuclear cooperation with proton exchange}

The Coulomb repulsion can be taken into account using an approximate form of
Coulomb-solution, which can be obtained from wave function describing
relative motion of like charges of charge numbers $z_{j}$ and $z_{k}$ \cite%
{Alder} and reads as $\varphi (\mathbf{r})=f_{jk}\left( E\right) e^{i\mathbf{%
k\cdot r}}/\sqrt{V}$ valid in the nuclear volume. Here $V$ denotes the
volume of normalization, $\mathbf{r}$ is the relative coordinate of the two
particles and $\mathbf{k}$ is the wave number vector in their relative
motion. $E$\ is the energy taken in the center of mass $\left( CM\right) $
coordinate system. $f_{jk}=\left( 2\pi \eta _{jk}/\left[ \exp \left( 2\pi
\eta _{jk}\right) -1\right] \right) ^{1/2}$ is the Coulomb factor and 
\begin{equation}
\eta _{jk}\left( E\right) =z_{j}z_{k}\alpha _{f}\sqrt{\frac{A_{j}A_{k}}{%
A_{j}+A_{k}}\frac{m_{0}c^{2}}{2E(CM)}}  \label{etajk}
\end{equation}%
is the Sommerfeld parameter, where $A_{j}$, $A_{k}$ are mass numbers of the
Coulomb interacting nuclei.

In the case of reaction $\left( \ref{R2}\right) $ momentum conservations ($%
\mathbf{k}_{n}=-\mathbf{k}_{3}$ in the final state and $\mathbf{k}_{p}=-%
\mathbf{k}_{n}(=$ $\mathbf{k}_{3})$ during $em$ interaction) furthermore
energy conservation $\hbar ^{2}\mathbf{k}_{n}^{2}/\left( 2m_{0}\right)
+\hbar ^{2}\mathbf{k}_{3}^{2}/\left( 2m_{0}A_{3}\right) =\Delta _{j}^{\pm }$
(in the final state) determine the (proton) energy of intermediate state $%
E_{p}\left( lab\right) $ in the laboratory frame of reference as $E_{p}(lab)=%
\left[ A_{3}/\left( 1+A_{3}\right) \right] \Delta _{j}^{\pm }$. Thus the
proton energy in the $CM$ system (of proton and nucleus of mass number $%
A_{2} $) $E_{p}(CM)=\frac{A_{2}}{1+A_{2}}E_{p}(lab)=\frac{A_{2}A_{3}}{\left(
1+A_{2}\right) (1+A_{3})}\Delta _{j}^{\pm }$ must be substituted in $\left( %
\ref{etajk}\right) $ that results 
\begin{equation}
\eta _{p2}\left( \xi _{j}^{\pm }\right) =z_{2}\alpha _{f}\sqrt{\frac{%
m_{0}c^{2}}{B\xi _{j}^{\pm }}}.  \label{etaCM}
\end{equation}

\subsection{Transition probability per unit time of spontaneous decay with
proton exchange}

The transition probability per unit time $W_{fi,\gamma p}^{-}$ of
spontaneous decay with proton exchange can be obtained with the aid of the
transition probability per unit time $W_{fi,\gamma n}^{-}$ of spontaneous
decay with neutron exchange modifying $I_{jn}^{-}$ in it as 
\begin{eqnarray}
I_{jp}^{-} &=&\int_{0}^{\delta _{j}}u^{3}f_{p2}^{2}(\xi _{j}^{\pm })
\label{Ijp-} \\
&&\times \left[ F_{jl_{f}}\left( \xi _{j}^{-}\right) \left[ S^{E}\left( \xi
_{j}^{-}\right) +\eta _{0}^{2}S^{M}\left( \xi _{j}^{-}\right) \right] \right]
du,  \notag
\end{eqnarray}%
where 
\begin{equation}
f_{p2}^{2}(\xi _{j}^{\pm })=2\pi \eta _{p2}\left( \xi _{j}^{\pm }\right)
/\left( \exp \left[ 2\pi \eta _{p2}\left( \xi _{j}^{\pm }\right) \right]
-1\right)  \label{Coulomb}
\end{equation}%
with $\eta _{p2}\left( \xi _{j}^{\pm }\right) $ and $\xi _{j}^{\pm }$ given
by $\left( \ref{etaCM}\right) $ and $\left( \ref{k0peralf}\right) $.

Furthermore, in estimating the cooperation length the $\Lambda \lesssim s%
\left[ E_{p}(CM)\right] $ choice (an upper estimate) seems to be acceptable
where $s\left[ E_{p}(CM)\right] $ is the stopping range of a proton of
energy $E_{p}(CM)$ determined above.

\subsection{Cross section of photo-induced nuclear cooperation with proton
exchange}

Similarly to the above, the cross section $\sigma _{\gamma p}^{\pm }$ of
photo-induced nuclear cooperation with proton exchange can be determined
from $\sigma _{\gamma n}^{\pm }$ as%
\begin{equation}
\sigma _{\gamma p}^{\pm }=\sum_{\varepsilon _{j}}K_{\sigma }uf_{p2}^{2}(\xi
_{j}^{\pm })F_{jl_{f}}\left( \xi _{j}^{\pm }\right) \left[ S^{E}\left( \xi
_{j}^{\pm }\right) +\eta _{0}^{2}S^{M}\left( \xi _{j}^{\pm }\right) \right] .
\label{sigmap}
\end{equation}%
In the rate of photo-induced nuclear cooperation with proton exchange the
quantity 
\begin{equation}
J_{jp}^{\pm }=\int_{u_{\min }}^{u_{\max }}uf_{p2}^{2}(\xi _{j}^{\pm })\left[
S^{E}\left( \xi _{j}^{\pm }\right) +\eta _{0}^{2}S^{M}\left( \xi _{j}^{\pm
}\right) \right] \left( \frac{d\Phi _{\gamma }}{du}\right) du  \label{Jjp+-}
\end{equation}%
must be used instead of $J_{jn}^{\pm }$ (se below).

\section{Explanation of observations}

\begin{table}[tbp]
\tabskip=8pt 
\centerline {\vbox{\halign{\strut $#$\hfil &\hfil$#$\hfil&\hfil$#$
\hfil&\hfil$#$\hfil\cr
\noalign{\hrule\vskip2pt\hrule\vskip2pt}
 Isotope&r_{A_{2}}&D_{0n}($MeV$)\cr
\noalign{\vskip2pt\hrule\vskip2pt}
Mo^{98}& 0.2413& 3.7009\cr
Mo^{100}& 0.0963 & 3.1739\cr
Er^{162}& 0.0014 & 5.2741\cr
Er^{164}& 0.0161 & 4.4255\cr
Er^{170}& 0.1493 & 3.4570\cr
Hf^{179}& 0.1362 & 5.1634\cr
Hf^{180}& 0.3508 & 3.4712\cr
\noalign{\vskip2pt\hrule\vskip2pt\hrule}}}}
\caption{Natural abundances $\left( r_{A_{2}}\right) $ and $D_{0n}=\Delta
_{d}-\Delta _{p}+\Delta _{A_{2}}-\Delta _{A_{2}+1}$values of those initial
isotopes, which are thought to be essential to the explanation of
observations of \protect\cite{steinetz}. Here $\Delta _{A_{2}}$, $\Delta
_{A_{2}+1}$, $\Delta _{d}$ and $\Delta _{p}$ are the energy excesses of
neutral atoms of mass numbers $A_{2}$, $A_{2}+1$, deuteron and proton,
respectively \protect\cite{Shir}. $D_{0n}$ is the energy of spontaneous
reaction $d+$ $_{z_{2}}^{A_{2}}X\rightarrow \protect\gamma +$ $p+$ $%
_{z_{2}}^{A_{2}+1}X$ into the ground state of the final nucleus of mass
number $A_{2}+1$.}
\end{table}

\subsection{Rate of isotope creation by photo-induced nuclear cooperation
with neutron exchange}

Now the rate of nuclear cooperation with neutron exchange is determined in a
photon-flux. The photon flux $d\Phi _{\gamma }$ in an energy interval $%
dE_{\gamma }$ can be written as $d\Phi _{\gamma }=\left( d\Phi _{\gamma
}/dE_{\gamma }\right) dE_{\gamma }$ where $\left( d\Phi _{\gamma
}/dE_{\gamma }\right) $ is the photon flux per unit photon energy. The rate $%
d\left[ dN_{n}/dt\right] $ due to $d\Phi _{\gamma }$ can be written as $d%
\left[ dN_{n}/dt\right] =N_{d}\left( d\Phi _{\gamma }/dE_{\gamma }\right)
dE_{\gamma }\left( \sigma _{\gamma n}^{+}+\sigma _{\gamma n}^{-}\right) $
and the full rate of nuclear events produced by photons in the energy range $%
E_{\gamma \min }<E_{\gamma }<E_{\gamma \max }$ can be written as%
\begin{equation}
\frac{dN_{n}}{dt}=N_{d}\int_{E_{\gamma \min }}^{E_{\gamma \max }}\left( 
\frac{d\Phi _{\gamma }}{dE_{\gamma }}\right) \sum_{\varepsilon _{j}}\left(
\sigma _{\gamma n}^{+}+\sigma _{\gamma n}^{-}\right) dE_{\gamma },
\label{dN/dtdef}
\end{equation}%
where $N_{d}$ is the number of deuterons in the volume $V_{s}$ and $%
N_{d}=V_{s}n_{d}$ with $n_{d}$ the deuteron number density $n_{d}$. Using
the variables $u=E_{\gamma }/B$, $\delta _{j}=\left( D_{0n}-\varepsilon
_{j}\right) /B$, $\eta _{0}^{2}=0.3684$ and $\xi _{j}^{\pm }=\left(
k_{0j}^{\pm }/\alpha \right) ^{2}=\left( \frac{A_{3}}{A_{3}+1}\right)
2\left( \delta _{j}\pm u\right) $ again the full rate has the form 
\begin{equation}
\frac{dN_{n}}{dt}=K_{N}\sum_{+,-,\varepsilon _{j}}J_{jn}^{\pm }
\label{dN/dt}
\end{equation}%
with $K_{N}=\Lambda n_{d}n_{z_{2}}r_{A_{2}}A_{3}C_{\sigma }$,%
\begin{equation}
J_{jn}^{\pm }=\int_{u_{\min }}^{u_{\max }}u\left[ S^{E}\left( \xi _{j}^{\pm
}\right) +\eta _{0}^{2}S^{M}\left( \xi _{j}^{\pm }\right) \right] \left( 
\frac{d\Phi _{\gamma }}{du}\right) du,  \label{Jj}
\end{equation}%
where $u_{\min }=E_{\gamma \min }/B$ and $u_{\max }=E_{\gamma \max }/B$. $%
E_{\gamma \min }$ and $E_{\gamma \max }$ are the lowest and highest possible
photon energies in the beam. Taking $\Lambda =V_{s}^{1/3}$%
\begin{equation}
K_{N}=V_{s}^{1/3}n_{d}n_{z_{2}}r_{A_{2}}A_{3}C_{\sigma }.  \label{KN2}
\end{equation}

The rate $dN_{n}/dt$ induced by irradiation of a photon flux is worth to
compare with the full spontaneous rate $W_{fi,\gamma n}^{-}$. Their ratio $%
\kappa $ is defined as $\kappa =\left( dN_{n}/dt\right) /W_{fi,\gamma n}^{-}$
and it is 
\begin{equation}
\kappa =\frac{C_{\sigma }}{C_{W}}\frac{\sum_{+,-,\varepsilon
_{j}}J_{jn}^{\pm }}{\sum_{\varepsilon _{j}}I_{jn}^{-}}  \label{kappa}
\end{equation}%
with $C_{\sigma }/C_{W}=\left( \hbar c\right) ^{3}\pi ^{2}/\left(
cB^{3}\right) =2.29\times 10^{-43}$ $cm^{2}s$. The order of magnitude of the
second fraction is determined by $d\Phi _{\gamma }/du=B\left( d\Phi _{\gamma
}/dE_{\gamma }\right) $ which varies in the range about $10^{12}$ $%
cm^{-2}s^{-1}<d\Phi _{\gamma }/du<10^{14}$ $cm^{-2}s^{-1}$ in the experiment
of \cite{steinetz} and therefore $\kappa $ can be estimated as $%
10^{-31}<\kappa <10^{-29}$. The order of magnitude estimation of $\kappa $
remains valid in the case of cooperative processes with proton exchange.
Consequently, one can conclude that $\gamma $ irradiation causes negligible
effect compared to the spontaneous process.

Investigating numerically the full transition probability per unit time ($%
W_{fi,\gamma n}^{-}$, see $\left( \ref{Wfign-}\right) $) of the spontaneous
process we take for example $n_{d}=n_{z_{2}}=a_{0}^{-3}$ with $a_{0}=4\times
10^{-8}$ $cm$, $\Lambda =1$ $cm$ and $A_{3}=100$ resulting $n_{d}\Lambda
n_{z_{2}}A_{3}C_{W}=196$ $s^{-1}$. As a model process the case of $d$ $-$ $%
^{98}Mo(0^{+})$ cooperation is taken in the Weisskopf-long wavelength
approximation (calculating $H_{jl_{f}}^{2}(\alpha R_{3}\sqrt{\xi _{j}^{\pm }}%
)$ with the aid of $\left( \ref{HknRA3LWA}\right) $). In this case the
contributions of levels of $l_{f}=0$ of $^{99}Mo\ $ ($1/2^{+}$ levels) are
taken into account. Their number of $\varepsilon _{j}<D_{0n}$ is 8. $%
\sum_{\varepsilon _{j}}I_{jn}^{-}=0.093$ and $r_{98}(Mo)=0.2413$ resulting $%
W_{fi,\gamma n}^{-}=4.4$ $s^{-1}$. (A somewhat smaller number is obtained in
the case of production of $^{100}Mo$. Natural abundances $\left(
r_{A_{2}}\right) $ and $D_{0n}=\Delta _{d}-\Delta _{p}+\Delta
_{A_{2}}-\Delta _{A_{2}+1}$ values of those initial isotopes, which are
thought to be essential to the explanation of observations of \cite{steinetz}
can be found in Table. I.)

The spontaneous process starts up as soon as the sample is made ready. In
the experiment of \cite{steinetz} there was $T=6$ $h$ irradiation, which can
be considered as a 'waiting time' from the point of view of the spontaneous
process. Thus at least $W_{fi,\gamma n}^{-}T\simeq 9.5\times 10^{4}$ nuclear
events happened during this time resulting $^{99}Mo$. Similarly in the cases
of the other initial isotopes (see Table I.) the spontaneous process yields
instable isotopes, and it is thought that their traces were observed by
gamma spectroscopy \cite{steinetz}.

\subsection{Neutron production}

In nuclear cooperation processes with proton exchange (see $\left( \ref{R2}%
\right) $) free neutrons are created. Since the Coulomb factor decreases
strongly with the increase of $z_{2}$ it is expected that neutrons are
created mainly in reactions with nuclei of small $z_{2}$. Considering the
compositions of samples in the experiment of \cite{steinetz}, the following
reactions may be candidates of source of neutron creation by cooperative
spontaneous $\gamma $ emission with proton exchange:

\begin{equation}
\text{ }d+d\rightarrow \gamma +n+\text{ }_{2}^{3}He+D_{0p},  \label{ddnHe}
\end{equation}%
\begin{equation}
d+\text{ }_{6}^{12}C\rightarrow \gamma +n+\text{ }_{6}^{13}N+D_{j},
\label{dCnN}
\end{equation}%
\begin{equation}
d+\text{ }_{6}^{13}C\rightarrow \gamma +n+\text{ }_{6}^{14}N+D_{j}.
\label{dC13nN14}
\end{equation}

Naturally their counterparts, i.e. the cooperative spontaneous $\gamma $
emission with neutron exchange reactions 
\begin{equation}
d+d\rightarrow \gamma +\text{ }p+\text{ }t+D_{0n},  \label{ddtp}
\end{equation}%
\begin{equation}
d+\text{ }_{6}^{12}C\rightarrow \gamma +p+\text{ }_{6}^{13}C+D_{j},
\label{dCpC}
\end{equation}%
\begin{equation}
d+\text{ }_{6}^{13}C\rightarrow \gamma +p+\text{ }_{6}^{14}C+D_{j},
\label{dC13pC14}
\end{equation}%
are possible too. However the direct observation of creation of $t$, $%
_{6}^{13}C$ and $_{6}^{14}C$ is rather hard.

\section{Conclusions}

The cross section of photo-induced nuclear cooperation and the transition
probabaility per unit time of cooperative spontaneous $\gamma $ emission
both with neutron and proton exchange are determined in deuterated
materials. It is found that the full rate of cooperative spontaneous $\gamma 
$ emission is many orders of magnitude larger than the rate of photo-induced
nuclear cooperation that would produce a $\gamma $ source of flux available
todate. It is found that the observed activity can not be achieved by
irradiation of samples by $\gamma $ flux. Perhaps, cooperative spontaneous $%
\gamma $ emission may be responsible for the observed nuclear activity. \cite%
{steinetz}.

\section{Appendix}

\subsection{Interaction Hamiltonians}

The electric and magnetic fields $\mathbf{E=}$ $\mathbf{e}E$ and $\mathbf{H}=%
\mathbf{e}_{k}\times \mathbf{e}E$ in electromagnetic wave are perpendicular
to each other and to the direction of propagation $\mathbf{e}_{k}=\mathbf{k}%
/\left\vert \mathbf{k}\right\vert $, where $\mathbf{E=}$ $\mathbf{e}E$ is
the electric field vector of the quantized field with $E=i\sum_{\mathbf{k},%
\mathbf{e}}\left( 2\pi E_{\gamma }/V_{\gamma }\right) ^{1/2}[ae^{-i\left(
\omega t-\mathbf{k\cdot r}\right) }-a^{+}e^{i\left( \omega t-\mathbf{k\cdot r%
}\right) }]$. Here $E_{\gamma }/\hbar =\omega $, $\mathbf{k}$ and $\mathbf{e}
$ are the angular frequency, wave number vector and vector of state of
polarization of $\mathbf{E}$, $V_{\gamma }$ is the volume of normalization, $%
a$ and $a^{+}$ are the photon annihilation and creation operators of the
quantized field. The energy of a photon of angular frequency is $E_{\gamma
}=\hbar \omega $.

The electric $\left( V_{E}^{em}\right) $ and magnetic $\left(
V_{M}^{em}\right) $ dipole interaction with electromagnetic radiation reads
in the electric dipole gauge and in the long wavelength (dipole)
approximation ($c/\omega \gg R_{A_{2}}$) as:%
\begin{equation}
V_{E}^{em}=-q\mathbf{r\cdot e}E_{0}\text{ and }V_{M}^{em}=-\mathbf{m\cdot }%
\left( \mathbf{e}_{k}\times \mathbf{e}\right) E_{0}  \label{Vem}
\end{equation}%
where $\mathbf{r}$ is the space vector of the particle having electric
charge $q$ and%
\begin{equation}
\mathbf{m}=\frac{e\hbar }{2m_{0}c}\left( \mu _{n}\mathbf{\sigma }_{n}+\mu
_{p}\mathbf{\sigma }_{p}\right)  \label{m-vector}
\end{equation}%
is the magnetic dipole operator with $\mu _{n}=-1.91$, $\mu _{p}=2.79$. $%
\mathbf{\sigma }_{n}$ and $\mathbf{\sigma }_{p}$ are vectors made from
Pauli-spinors (the indices $n$ and $p$ refer to neutron and proton,
respectively,) and $E_{0}=i\sum_{\mathbf{k},\mathbf{e}}\left( 2\pi E_{\gamma
}/V_{\gamma }\right) ^{1/2}[ae^{-i\omega t}-a^{+}e^{i\omega t}]$.

For the strong interaction the interaction potential%
\begin{equation}
V^{st}\left( \mathbf{x}\right) =-V_{0}\text{ \ if }\left\vert \mathbf{x}%
\right\vert \leq b\text{ and }V^{st}\left( \mathbf{x}\right) =0\text{ \ if }%
\left\vert \mathbf{x}\right\vert >b  \label{VSt1}
\end{equation}%
is applied, where the choice for $V_{0}=45$ $MeV$ and $b=r_{0}A_{2}^{1/3}$
with $r_{0}=1.2\times 10^{-13}$ $cm$ seem to be reasonable in the case of
target particle \cite{Blatt}, \cite{Pal}.

\subsection{Matrix elements of $V^{em}$}

The matrix element of the interaction potential of electromagnetic radiation
with matter between the initial and intermediate states $%
V_{ki}^{em}=V_{E,ki}^{em\pm }+V_{M,ki}^{em\pm }$ according to electric $%
\left( V_{E}^{em}\right) $ and magnetic $\left( V_{M}^{em}\right) $ dipole
interaction with electromagnetic radiation. The upper indices $+$ and $-$
correspond to absorption ($+$) and induced emission ($-$), respectively. 
\begin{equation}
V_{E,ki}^{em\pm }=g^{\pm }V_{E,ki,0}^{em}\sin \Theta _{\mathbf{k}_{p}}\frac{%
(2\pi )^{3}}{\sqrt{V_{n}V_{p}}}\delta \left( \mathbf{k}_{n}+\mathbf{k}%
_{p}\right) ,  \label{VemEki}
\end{equation}%
where $\Theta _{p}$ is the angle between $\mathbf{k}$ of incident photon and 
$\mathbf{k}_{p}$, $g^{+}=\sqrt{n_{\gamma }}$ and $g^{-}=\sqrt{n_{\gamma }+1}$%
. 
\begin{equation}
V_{E,ki,0}^{em}=\frac{1}{V_{s}^{1/2}}\frac{ei\left( 2\pi E_{\gamma
}/V_{\gamma }\right) ^{1/2}}{2\left( 1-\alpha r_{t}\right) ^{1/2}}\frac{ik%
\sqrt{8\pi \alpha }}{\left( \alpha ^{2}+k^{2}\right) ^{2}},  \label{VemEki0}
\end{equation}%
with $k=\left\vert \mathbf{k}_{p}-\mathbf{k}_{n}\right\vert /2$. The factor $%
\left( 1-\alpha r_{t}\right) ^{-1/2}$ comes from range correction of the
zero range approximation of nuclear force \cite{Blatt} with $%
r_{t}=1.759\times 10^{-13}$ $cm$ \thinspace\ \cite{Arenhovel}.%
\begin{equation}
V_{M,ki}^{em\pm }=g^{\pm }\eta V_{E,ki,0}^{em}\frac{(2\pi )^{3}}{\sqrt{%
V_{n}V_{p}}}\delta \left( \mathbf{k}_{n}+\mathbf{k}_{p}\right) 
\label{VemMki}
\end{equation}%
with $\eta =[\frac{2}{3}(\sigma ^{MD}/\sigma ^{ED})]^{1/2}$, where $\sigma
^{MD}$ and $\sigma ^{ED}$ are the magnetic and electric dipole parts of
regular photodissociation cross section \cite{Blatt}, \cite{Arenhovel}.
Taking $\sigma ^{MD}$ and $\sigma ^{ED}$ from \cite{Blatt} $\eta =\eta
_{0}\chi \left( k\right) $ with%
\begin{equation}
\eta _{0}=\left\vert \mu _{n}-\mu _{p}\right\vert \left\vert 1-\alpha
a_{s}\right\vert \sqrt{B/\left( 6m_{0}c^{2}\right) }  \label{eta0}
\end{equation}%
and%
\begin{equation}
\chi \left( k\right) =\left[ \left( \frac{1+\left( \frac{k}{\alpha }\right)
^{2}}{k/\alpha }\right) ^{2}\frac{1}{\left( 1+\left( \frac{k}{\alpha }%
\right) ^{2}\alpha ^{2}a_{s}^{2}\right) }\right] ^{1/2}.  \label{khik}
\end{equation}%
Here $a_{s}$ is the scattering length in the singlet state. Taking $%
a_{s}=-2.37\times 10^{-12}$ $cm$ \cite{Blatt} $\eta _{0}=0.607$.

\subsection{Matrix elements of $V^{st}$ - Cooperation}

$V_{fk}^{st}$ is the matrix element of the potential of the strong
interaction between the intermediate and final states. When calculating $%
V_{fk}^{st}$ it must be taken into account that the cooperating nuclei are
located at a distance $L$ far from each other. The amplitude $A\left(
L\right) $ of the emitted neutron spherical wave in the $kL\rightarrow
\infty $ limit varies as $\left\vert A(L)\right\vert =\sin \left(
k_{n}L\right) /\left( k_{n}L\right) $ (e.g. in an $s$-wave) or $\left\vert
A(L)\right\vert =\cos \left( k_{n}L\right) /\left( k_{n}L\right) $ (e.g. in
a $p$-wave). Since $L$ is very large compared to nuclear extension the wave
appearing at the cooperating (neutron absorbing) nucleus may be considered
to be a plane wave of form $A(L)e^{ikz}$ with an appropriate choice of the
frame of reference. With the aid of a state of this type%
\begin{eqnarray}
V_{fk}^{st} &=&-V_{0}R_{3}^{3/2}i^{l_{f}}\sqrt{12\pi \left( 2l_{f}+1\right) }
\label{Vstfk} \\
&&\times H_{jl_{f}}\left( k_{n}R_{3}\right) \frac{\sin \left( k_{n}L\right) 
}{\left( k_{n}L\right) }\frac{\left( 2\pi \right) ^{3}}{\sqrt{V_{n}V_{t}V_{3}%
}}\delta \left( \mathbf{k}_{n}-\mathbf{k}_{3}\right)  \notag
\end{eqnarray}%
in the case of $A(L)=\sin \left( k_{n}L\right) /\left( k_{n}L\right) $ and%
\begin{equation}
H_{jl_{f}}\left( k_{n}R_{3}\right) =\int_{0}^{1}\phi _{j}\left(
R_{3}x\right) j_{_{l_{f}}}\left( k_{n}R_{3}x\right) x^{2}dx.  \label{HknRA3}
\end{equation}%
In the Weisskopf-approximation%
\begin{equation}
H_{jl_{f}}^{W}\left( k_{n}R_{3}\right) =\int_{0}^{1}j_{_{l_{f}}}\left(
k_{n}R_{3}x\right) x^{2}dx.  \label{HknRA3W}
\end{equation}%
In the long wavelength-approximation (LWA, $k_{n}R_{3}\ll 1$ case) $%
j_{_{l_{f}}}\left( k_{n}R_{3}x\right) =\left( k_{n}R_{3}x\right)
^{l_{f}}/\left( 2l_{f}+1\right) !!$ which gives (in the
Weisskopf-approximation)%
\begin{equation}
H_{jl_{f}}^{W}\left( k_{n}R_{3}\right) =\frac{\left( k_{n}R_{3}\right)
^{l_{f}}}{\left( l_{f}+3\right) \left( 2l_{f}+1\right) !!}.
\label{HknRA3LWA}
\end{equation}%
The case $l_{f}=0$ gives the leading term with $H_{jl_{f}}^{W}\left(
k_{n}R_{3}\right) =1/3$ in the LWA.

\subsection{Some details of calculation of $W_{fi,\protect\gamma n}$}

In the center of mass frame of reference momentum conservation leads to the
appearance of wave number vector Dirac-deltas $\delta \left( \mathbf{k}_{n}+%
\mathbf{k}_{p}\right) $ and $\delta \left( \mathbf{k}_{n}-\mathbf{k}%
_{3}\right) $). Integrating first over $\mathbf{k}_{n}$ it results $\mathbf{k%
}_{n}=-\mathbf{k}_{p}$ substitution in the integrand of $\int
V_{fk}^{st}V_{ki}^{em}\Delta E_{ki}^{-1}\delta \left( \mathbf{k}_{n}+\mathbf{%
k}_{p}\right) \delta \left( \mathbf{k}_{n}-\mathbf{k}_{3}\right)
d^{3}k_{n}\left( 2\pi \right) ^{-3}V_{n}=T_{fi}(\mathbf{k}_{3},\mathbf{k}%
_{p})$ while the volume of normalization $V_{n}$ of the neutron disappears.
The square of the remaining Dirac delta $\delta ^{2}\left( \mathbf{k}_{p}+%
\mathbf{k}_{3}\right) =\delta \left( 0\right) \delta \left( \mathbf{k}_{p}+%
\mathbf{k}_{3}\right) =V_{3}\left( 2\pi \right) ^{-3}\delta \left( \mathbf{k}%
_{p}+\mathbf{k}_{3}\right) $. Taking $V_{3}=V_{t}$, $V_{t}$ disappears too.
The factors $V_{3}\left( 2\pi \right) ^{-3}$ and $V_{p}\left( 2\pi \right)
^{-3}$ coming from phase space factors of proton and $_{z_{2}}^{A_{2}+1}X$
final nucei make disappearing $V_{3}$ and $V_{p}$ too. Then integrating $%
\left\vert T_{fi}(\mathbf{k}_{3},\mathbf{k}_{p})\right\vert ^{2}$ over $%
\mathbf{k}_{3}$ gives $\mathbf{k}_{3}=-\mathbf{k}_{p}$ substitution in the
standard $W_{fi,\gamma n}$ calculation.

The energy denominator $\Delta E_{ki}=E_{k}-E_{i}-\Delta _{ik}\mp E_{\gamma
} $, where $\Delta _{ik}=\Delta _{-}-\Delta _{n}=-B$~is the difference
between the rest energies of the initial and intermediate states. $\Delta
_{-}=\Delta _{d}-\Delta _{p}$, $-E_{\gamma }$ and $E_{\gamma }$ correspond
to photon absorption and emission, $E_{i}$, $E_{k}$ and $E_{f}$ are the
kinetic energies in the initial, intermediate and final states,
respectively, $E_{i}=0$ is supposed.

The energy denominator $\left( \Delta E_{ki}\right) $ reads as 
\begin{equation}
\Delta E_{ki}=\frac{\hbar ^{2}\mathbf{k}_{p}^{2}}{m_{0}}+B\mp E_{\gamma }
\label{DeltaEmui2}
\end{equation}%
after the substitution $\mathbf{k}_{n}=-\mathbf{k}_{p}$. Using $\mathbf{k}%
_{3}=-\mathbf{k}_{p}$, the final kinetic energy $E_{f}$ in the argument of
energy Dirac-delta $\left[ \delta \left( E_{f}-\Delta _{j}^{\pm }\right) %
\right] $ is%
\begin{equation}
E_{f}=\frac{\left( A_{3}+1\right) \hbar ^{2}\mathbf{k}_{p}^{2}}{2A_{3}m_{0}}.
\label{Ef22}
\end{equation}%
The energy Dirac-delta is converted into $\delta \left( k_{0j}^{\pm
}-k_{p}\right) $ where $k_{p}=\left\vert \mathbf{k}_{p}\right\vert $ and $%
k_{0j}^{\pm }=\hbar ^{-1}\sqrt{2m_{0}\left[ A_{3}/\left( A_{3}+1\right) %
\right] \Delta _{j}^{\pm }}$.

\end{document}